\documentclass[useAMS,usenatbib]{mn2e}

\newcommand{\etal}{\emph{et al.}}
\usepackage{graphicx}       		
\graphicspath{{./}{Figs_new/}}
\usepackage{subfigure}
\usepackage{epstopdf}
\usepackage[breaklinks=true]{hyperref}
\usepackage{mathtools}   
\usepackage{color}
\definecolor{Blue}{rgb}{0,0,0.9}

\newcommand*{\bfrac}[2]{\genfrac{}{}{0pt}{}{#1}{#2}}

\title[Mutual approximations: application to the Galilean moons.]{Astrometry of mutual approximations between natural satellites. Application to the Galilean moons.\thanks{Based on observations made at the Laborat\'orio Nacional de Astrof\'isica
(LNA), Itajub\'a-MG, Brazil.}
}
\author[B. Morgado \etal]{B. Morgado$^{1,2}$\thanks{E-mail:
Morgado.fis@gmail.com}
, M. Assafin$^{2}$, R. Vieira-Martins$^{1,2}$, J.I.B. Camargo$^{1}$,\newauthor 
 A. Dias-Oliveira$^{1}$, A. R. Gomes-J\'unior$^{2}$\\
$^{1}$Observat\'orio Nacional/MCTI, R. General Jos\'e Cristino 77, Rio de Janeiro – RJ 20.921-400, Brazil\\
$^{2}$Observat\'orio do Valongo/UFRJ, Ladeira Pedro Antonio 43, Rio de Janeiro – RJ 20080-090, Brazil}

\begin{document}

\date{Accepted 2016 May 19. Received 2016 May 18; in original form 2016 April 08}

\pagerange{\pageref{firstpage}--\pageref{lastpage}} \pubyear{2015}

\maketitle

\label{firstpage}

\begin{abstract}

Typically we can deliver astrometric positions of natural satellites with errors in the 50-150 mas range. Apparent distances from mutual phenomena, have much smaller errors, less than 10 mas. However, this method can only be applied during the equinox of the planets. We developed a method that can provide accurate astrometric data for natural satellites -- the mutual approximations. The method can be applied when any two satellites pass close by each other in the apparent sky plane. The fundamental parameter is the central instant $t_0$ of the passage when the distances reach a minimum.

We applied the method for the Galilean moons. All observations were made with a 0.6 m telescope with a narrow-band filter centred at 889 nm with width of 15 nm which attenuated Jupiter's scattered light. We obtained central instants for 14 mutual approximations observed in 2014-2015. We determined $t_0$ with an average precision of 3.42 mas (10.43 km). For comparison, we also applied the method for 5 occultations in the 2009 mutual phenomena campaign and for 22 occultations in the 2014-2015 campaign. The comparisons of $t_0$ determined by our method with the results from mutual phenomena show an agreement by less than 1-sigma error in $t_0$, typically less than 10 mas. This new method is particularly suitable for observations by small telescopes.  

\end{abstract}


\section{Introduction} \label{int}

The formation of planets occurs in a disk of gas and dust around the protostar. The formation of regular satellites around a giant planet probably occurs in a similar way in a circum-planetary disk around the planet. However, there are different models for this formation in the literature with pros and cons each \citep{Crida_2012}. The orbital evolution of these regular satellites around a giant planet can give us hints about their formation.
 
The orbital studies of these celestial bodies demand the observation of positions, relative distances or other forms of observables, like central instants and impact parameters in mutual phenomena, over extended period of time to fit them with dynamic models \citep{Sitter1928,Lieske1987,Lainey2009}. However, to detect very weak disturbance forces, such as \emph{tidal forces}, these observables have to be very precise.
This high precision usually cannot be attained with classical methods. For example, Jupiter and Saturn's brightness makes it very difficult to obtain enough number of astrometric calibration stars in a CCD frame in order to obtain a good individual position for its main satellites. The classical CCD astrometry of a single satellite furnishes positions with precisions (usually the standard deviation from a set of a few hundreds of images) in the range of 50 to 150 mas \citep{Stone2001,Kiseleva2008}. This issue was partially solved using relative positions between two (or more) satellites \citep{Veillet1980,Veiga1987,Veiga1994,Harper1997,Vienne2001,Peng2008}. More recently \cite{Peng2012} achieved relative positions with precision of 30 mas.   

Some of these works use the ephemeris of the satellites as a ruler in order to obtain the reference system orientation and scale in the CCD image. Often, one uses the positions of bodies with more precise ephemeris to determine the position of another body with a worst ephemeris. An example is the case of Miranda of Uranus, when the ephemeris of the others satellites (mainly Oberon) were used as reference frame in the position reductions \citep{Veillet1980}. Another possibility is the use of the so called \emph{precision premium} \citep{Peng2008}, first pointed out by \cite{Pascu1994}. The idea is that the accuracy of the determination of apparent distances is remarkably improved for short apparent distances (less than 85"), as instrumental and astronomical issues tend to affect the images of both satellites in the same way. In principle, we can use the ephemeris distance between two small-separated satellites and the actual observed distance to determine the scale and orientation of the image. This method was mainly used for the Saturnian main satellites \citep{Harper1997,Vienne2001} and for the Galilean satellites \citep{Peng2012}.
 
Another way to obtain very precise relative positions of typically a few mas is with the observation of mutual phenomena. However this can only be done during the equinox of these planets -- see for example \cite{ Assafin2009,Emelyanov2009,Emelyanov2011,Diasoliveira2013,Arlot2014}. Mutual phenomena consist of occultations and eclipses between natural satellites, as seen along the line of sight of an observer. One satellite (or its projected shadow) hides the other causing a drop in the light flux which can be measured with high precision using differential photometry in a small FOV, when astronomical and instrumental systematic errors in the measurements of the targets and calibration sources tend to cancel out. This flux variation is connected to the relative apparent motion of the satellites in the sky, and thus, ultimately, to their orbits. The apparent relative motion between the satellites under mutual phenomena can be described in terms of the instant of minimum apparent distance (central instant of the event), the minimum apparent distance at this instant (impact parameter), and the relative apparent velocity between one satellite (or shadow) and the other (\cite{Emelyanov2003, Diasoliveira2013} and references therein). The particular nature of orbital geometry in mutual phenomena offers valuable constrains in the orbital solutions of the satellites \citep{Lainey2004}. However, these phenomena can be observed only every 6 years for Jupiter, every 15 years for Saturn and every 42 years for Uranus \citep{Arlot2012,Arlot2013,Arlot2014}.    

Inspired in the mutual phenomena and based in the same orbital geometry, we propose the observation of what we call \emph{mutual approximations} between natural satellites in order to obtain the central instant when the apparent distance between two satellites reaches a minimum.

In mutual approximations the bodies only approach each other in the sky plane at adequate, not too short, distances -- we actually avoid occultations, we wish to be able to measure the satellites separately. Similarly to the geometric parameters of mutual phenomena, we can also determine the impact parameter and the relative velocity, but, in this case, only in pixel units. For orbit fitting, these supplementary parameters can only be useful  if we can accurately convert the distances from pixels to arc seconds using a standard reference frame, like an astrometric star catalogue representative of the ICRS. However, this is frequently not the case, as usually an insufficient number of reference stars is available in the FOV, for two reasons: short FOV and/or short exposure times to avoid saturation of the satellite images and light scatter from the central planet. One way to overcome this problem is to observe a crowded star field nearby and determine the pixel scale to be used. Another alternative is to use an ephemeris of the pair of satellites as the reference frame. But, by doing that we scale the observed impact parameter and the relative velocity to the used ephemeris, so no trully independent new results are really obtained, although they may serve to check parameters for internal consistency (see more details and computations in this respect in Section \ref{Anexo_A}).

In fact, \cite{Arlot_1982} were the first to suggest such astrometric approach. Also, \cite{Mason_1999} reported observations which they called \emph{"close pairings"} of the Galilean moons with a speckle interferometer mounted on the 26-inch refractor at the Naval Observatory in Washington, DC. But no further references, results or studies in this subject were published in the literature to our knowledge.

We  give  in  this  paper a complete description of the method of mutual approximations and present the results of its application to an observational campaign of mutual approximations for the Galilean moons carried out in 2014-2015. Our observations were made with the 0.6 m Zeiss telescope of the Observat\'orio do Pico dos Dias (OPD, Brazil) with a narrow-band methane filter centred at 889 nm with width of 15 nm that attenuates Jupiter's scattered light.

The paper is arranged as follows. In Section \ref{model}, we describe the theoretical model of relative motion for the mutual approximations and how to obtain the central instant of the event.  In Section \ref{Anexo_A} we show also how to compute the impact parameter and relative velocity. In Section \ref{reduc} we discuss what astronomical effects like solar phase-angle, atmospheric refraction and aberration could cause an offset in the raw measured central instant. In Section \ref{results} we detail the observational campaign of the mutual approximations for the Galilean satellites, the measurement of the images and our results. In Section \ref{comparar} we compare the results from mutual approximations and mutual phenomena simultaneously observed during the mutual phenomena campaign of 2009. For future comparisons, we also give the results of the approximations observed during the 2014-2015 equinox of Jupiter. Discussion and conclusions are presented in Section \ref{conclusao}.

\section{Apparent relative motion of satellites in mutual approximations. Determination of the central instant} \label{model}

We set here a geometric model that describes the variation with time of the apparent distance $d$ in the plane of the sky between two satellites for a short arc of their orbits, typical of mutual approximations. We assume that there is no strong deformation from the telescope optics on the images. A simple description by a polynomial in $t$ of arbitrary degree $n$ (typically $n=4$) is obtained using  $d^2$. In  mutual  approximations,  $d$  always  reaches  a minimum and the polynomial curve will present a positive concavity.

The same geometric parameters that describe mutual phenomena also describe mutual approximations, i.e. central instant, impact parameter and relative velocity. Notice however that only the central instant can be determined independently of any reference frame, even without the knowledge of the pixel scale of the images. But astronomical corrections such as refraction, aberration and solar phase angle must be taken into account (see Sect. \ref{reduc}). However, if one does want to get all the parameters, one needs to convert distances from pixels to arcseconds. In this case, the pixel scale must be computed. (see Sect. \ref{Anexo_A}).

The vector distance between the two satellites in the sky plane $\Delta \vec{r}(t)$ in a specific instant $t$ is:

\begin{equation}
\Delta \vec{r}(t) = \vec{r}_{1}(t) - \vec{r}_{2}(t)\label{eq:Ad3}
\end{equation}

Assuming that the square of the distance between two satellites in the short arc of their orbits, nearby the central instant of the approximation, can be described as a polinonm of arbitrary power $n$ in time, then, considering (\ref{eq:Ad3}) we obtain:

\begin{eqnarray}
d^{2}(t) &=& \Delta\vec{r}(t).\Delta\vec{r}(t) \nonumber\\
d^{2}(t) &=& a_{0} + a_{1}t + a_{2}t^{2} +...+ a_{n}t^{n} \label{eq:Ad4}
\end{eqnarray}
where the parameters $a_0$, $a_1$, $a_{n}$ are related with the kinematics of the satellites' motion.

Using least squares we fit this model with observed distances. Alternatively, we can fit ephemeris' apparent distances to study the event and even for testing the most adequate polynomial power to use in the fitting.

The central instant $t_0$ of the mutual approximation is obtained when the first time derivative of equation (\ref{eq:Ad4}) is set equal to zero.

\begin{equation}
a_{1} + 2a_{2}t_{0} + 3a_{3}t_{0}^{2} + .. + na_{n}t_{0}^{n-1} = 0 \label{eq:At1}
\end{equation}

The central instant $t_0$ comes from the determination of the root of Eq. (\ref{eq:At1}). Using an interactive process we determine the central instant, subtract it in $t$  and then redo the adjustment in order to set $t_0 = 0$. The central instant error comes from around-zero derivative of Eq. (\ref{eq:At1}) as a function of the error of the other coefficients, computed in the least squares fit. This error is given by Eq. (\ref{eq:At3}):

\begin{equation}
 \delta t_0 = - \frac{\delta a_1}{2a_2} \label{eq:At3}
\end{equation}

Note that the numerator of Eq. (\ref{eq:At3}) relates to the observed distance errors, thus it reflects as expected the noise in the observed apparent distance curve, related with the atmospheric conditions of the night. The error of the central instant is also dependent on the curve concavity related to $a_2$, which in turn is dominated by the relative velocity. A smaller central instant error is expected from a more pronounced minimum which comes from larger $a_2$ values, or higher relative velocities.

In practice the observational procedure in mutual approximations consists in observing a number of images, an hour before and after a central instant foreseen from an ephemeris. We then measure the apparent distances and, with the help of any ephemeris, apply corrections due to solar phase angle, refraction and aberration that could shift the distances in a way that could offset the central instant (see Sect. \ref{reduc}). The instant errors can be translated from seconds of time to mas or km by using the relative velocity and distance to the observer given by an ephemeris. In Section \ref{Anexo_A} we describe the procedure in the case that one wants to get the complete set of parameters, that is the impact parameter and relative velocity too, in the same fashion as in mutual phenomena.

How well does the model of apparent relative motion introduced here really describe the actual relative apparent path between two real satellites in a mutual approximation? Considering that the Galilean moons of Jupiter are amongst the most studied dynamical systems, and taking their most precise ephemeris up to date as representative of their real paths in the sky, we addressed this question by fitting our model directly to apparent distances computed from their ephemeris. In this work all events were well fitted by a fourth degree polynomial.  

\begin{figure}
\includegraphics[height=07cm]{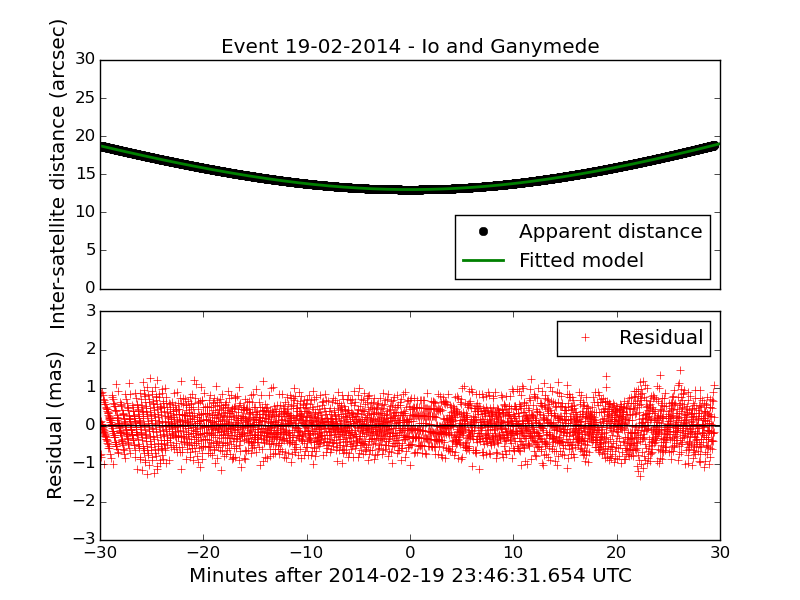}
\caption{Apparent sky distances between Io and Ganymede in the mutual approximation of February, 19th 2014. We used the NOE-5-2010-GAL ephemeris obtained in its topocentric form for OPD. Ephemeris distances $d$ are in black and model fitted ones in green. We took ephemeris positions every second. Notice that the fit is actually done in $d^2$, using Eq. (\ref{eq:Ad4}).}
\label{Fig_MA1}
\end{figure}

We chose as an example the approximation between Io and Ganymede at February 19th, 2014. The ephemeris used was the NOE-5-2010-GAL provided by IMCCE \footnote{Institut de M\'ecanique C\'eleste et de Calcul des \'Eph\'em\'erides;\\ Website: \url{http://www.imcce.fr/}} from \cite{Lainey2009}. In the Figure \ref{Fig_MA1} we computed the apparent distances from the ephemeris and fitted them with a fourth degree polynomial. The comparison between the fitted (green line) and ephemeris (black dots) apparent distances. The residual dispersion is also illustrated in the bottom (red crosses), in the sense \emph{model minus ephemeris}. The time resolution of the ephemeris was one second.

\begin{figure}
\includegraphics[height=07cm]{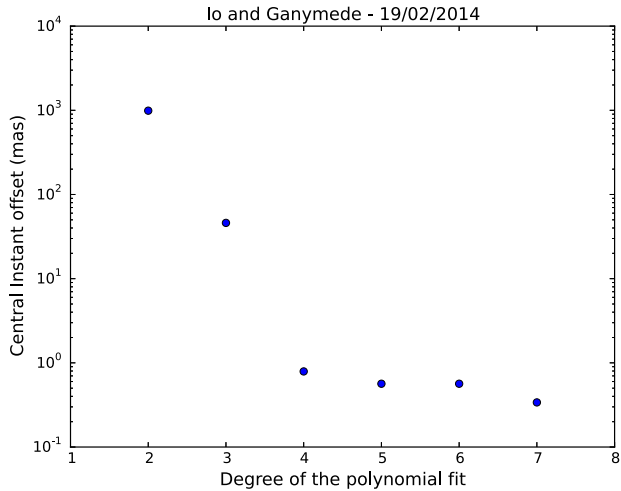}
\caption{Central instant offset for each polynomial
fitting. After the fourth degree there is no significant improvement in accuracy.}
\label{Fig_MA2}
\end{figure}

As seen in Figure \ref{Fig_MA1}, our model fits quite well the satellite ephemeris. The residual of the adjustment stays in 1 mas, corresponding to the numerical limit of extracted ephemeris positions, here truncated for computational purposes (notice that the actual ephemeris precision is worse than this). In Figure \ref{Fig_MA2} we illustrate the discrepancy between the value for the central instant in mas obtained using different polynomial models in time for $d^{2}$. Models with power less than the fourth degree give incomplete descriptions of the satellite's relative motion. We show that after the fourth degree there is no significant improvement in the precision or accuracy of the model, for this example. Others tests showed similar results.  

\section{Obtaining the impact parameter and relative velocity in mutual approximation} \label{Anexo_A}

Although the determination of the central instant comes from fits to the raw measured apparent distances given in pixel units (see Sect. \ref{model}), the UTC time is recorded in the observations with a GPS system or precise internet timing. Thus, the central instant derived in UTC is the only truly independent parameter obtained in mutual approximations without regard to any reference frame, be it a star catalogue or an ephemeris. For this reason, the central instant is the main parameter derived from the mutual approximations.

The impact parameter and relative velocity are also derived from fits to the observed apparent distances, but in this case they share the same metric. Thus, these parameters can only be obtained in standard units if the observed distances are also given in standard metrics, such as arcseconds. A conversion from pixels to arcseconds must be made. This can only be accomplished with the use of a reference frame.

Notice that if we use a star catalogue as reference frame, the obtained impact parameter and relative velocity are independent from any ephemeris. In this case, like in mutual phenomena, they may be as useful as the central instant in orbit fit work. This is possible if we have enough catalogue reference stars and can make the right ascension and declination reduction of the FOV. If this is not the case, we can alternatively reduce a nearby field with a sufficient number of reference stars and use the derived pixel scale in the FOV of the satellites to compute the observed distances. Unfortunately, this is rarely the case of Jupiter and Saturn. They are currently crossing a sky path with few stars. Besides, due to the brightness of their main moons and their proximity to the planet, the very short exposure times used prevent the imaging of reference stars. But Uranus may be a good and promising example for the use of this procedure.  

We can also use an ephemeris to obtain the pixel scale, see \cite{Peng2008}. This is very useful when we have no reference stars, or even when only the two satellites are available in the FOV. This is frequently the case of the observations of Jupiter and Saturn moons. The drawback is that the observed distances are scaled according to the ephemeris frame, so that the impact parameter and the relative velocity will be dependent of the used ephemeris to some extent, and may be not useful in orbital works. Even so, it is always a good practical procedure to compute the impact parameter and relative velocity using the ephemeris as reference, for it serves as an extra checking procedure of the results. For this reason, we describe this procedure in Sect. \ref{ps}.

\subsection{Modelling the impact parameter and relative velocity}\label{A2}

Since the observed satellite’s images are affected by solar phase angle, aberration and refraction, these effects are taken into account prior to fitting the observed distances with the topocentric ephemeris ones (see Sect. \ref{reduc}).

From Eq. (\ref{eq:Ad4}), when $t_0 = 0$, we see that the minimum distance between the satellites, the impact parameters $d_0$ of the approximation, will be related with $a_0$ in the form of
\begin{equation}
d_0 = \sqrt{a_0}
\end{equation}
and the incertitude of this parameter is
\begin{equation}
\delta d_0 = \frac{\delta a_0}{2d_0}
\end{equation}

The relative velocity is determined as follows: Using two consecutives images we determine the instantaneous variation of the relative position in x and y in the CCD frame. With the acurate time of each image we can determine the relative instantaneous velocity between both frames. Then we fit a linear function in the velocity curve in time obtained with all the images. The relative velocity of interest is the one for the central instant $t_0$ and its error is obtained from the linear fit.

Notice that, from our fit, these parameters are in pixel units and pixels per second. The conversion to arcsecond units is explained in the following Section.
 
\subsection{Determining the pixel scale} \label{ps}

Knowing from an ephemeris the theoretical apparent distances between the satellites ($\Delta\alpha\cos\delta,~ \Delta\delta$) in arcseconds, affected by solar phase angle, aberration and atmospheric refraction, and the instrumental distances ($\Delta x,~ \Delta y$) in pixels, makes it possible to obtain the pixel scale. Notice that we projected the ephemeris on the tangent plane using the gnomonic projection, so that the standard coordinates (X,Y) are used in the computation of the pixel scale.

We compute the pixel scale $P_{s}$ as the slope of a linear function fitted to the ratio between the ephemeris distances $d_e$ and instrumental distances $d_o$ (Eq. \ref{eq:ps}).

\begin{equation}
P_s = \frac{d_e}{d_o} \label{eq:ps}
\end{equation}

\begin{figure}
\includegraphics[height=07cm]{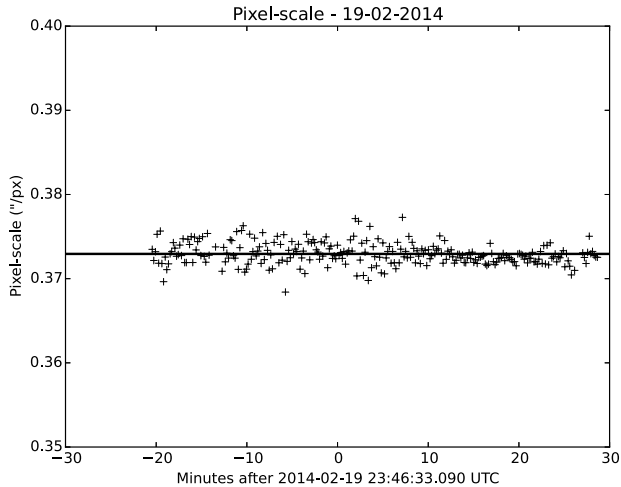}
\caption{Pixel-scale determined for the approximation between Io and Ganymede of February, 19th 2014 in arcseconds per pixel as a function of the time. We used the NOE-5-2010-GAL ephemeris obtained in its topocentric form for OPD.}
\label{Fig_ps}
\end{figure}

Figure \ref{Fig_ps} illustrates the variation of the pixel scale overnight for the mutual approximation between Io and Ganymede which occurred at February, 19th, 2014. For illustration, in this example we computed an error of $0.9~mas.pixel^{-1}$ for the pixel scale.

For comparison, we determined all the parameters of a mutual approximation, central instant, impact parameter, relative velocity and the pixel-scale using two different ephemeris. We chose the NOE-5-2010-GAl provided by IMCCE from \cite{Lainey2009} and the JPL\footnote{Jet Propulsion Laboratory; \\Website: \url{http://ssd.jpl.nasa.gov/}} ephemeris, $jup310$, for this analysis. The results and errors are displayed in Table \ref{tb:compar} for the approximation between Io and Ganymede of February, 19th of 2014. Other examples showed similar results.  

\begin{table}
\caption{Comparison between the results for the central instant, $t_{0}$, impact parameter, $d_{0}$ and relative velocity, $v_{r}$ for Io and Ganymede mutual approximation of February, 19th 2014 and the calibration parameter, pixel-scale $P_{s}$, using two ephemeris -- NOE-5-2010-GAL and $jup310$.}
\begin{tabular}{ccc}
\hline
Parameters &\multicolumn{2}{c|}{NOE-5-2010-GAL}  \\
 & Observation & Ephemeris  \\
\hline
$t_{0}~(hh:mm:ss)$ & 23:46:35.79 (0.76)& 23:46:31.68 (0.00)\\
$d_{0}~(mas)$ & 12982.89 (6.08)& 12995.12 (0.12)\\
$v_{r}~(mas/s)$ & 7.53 (0.08)& 7.54 (0.00)\\
$P_s~(''/px)$ & \multicolumn{2}{c|}{0.3729 (0.0009)}  \\
\hline
Parameters & \multicolumn{2}{|c}{JPL -- $jup310$} \\
& Observation & Ephemeris \\
\hline
$t_{0}~(hh:mm:ss)$ &  23:46:35.79 (0.76)& 23:46:35.22 (0.00)\\
$d_{0}~(mas)$ &  12974.24 (6.07)& 12996.66 (0.12)\\
$v_{r}~(mas/s)$ & 7.52 (0.08) & 7.58 (0.00)\\
$P_s~(''/px)$ &  \multicolumn{2}{|c}{0.3728 (0.0009)} \\
\hline
\label{tb:compar}
\end{tabular}
\end{table}

Notice that the values for the observational parameters are consistent and almost independent of the ephemeris utilized.

\section{Astronomical corrections}\label{reduc}

Astronomical effects such as the solar phase angle, the atmospheric refraction and aberration affect the apparent distance between the satellites. They must be taken into account when we want to obtain all the parameters, including the impact parameter and relative velocity, as described in Sect 3. But which of these effects can actually offset the central instant in a mutual approximation? And to what typical amounts? We address these question in this Section.

In the case of the solar phase angle, the photocentre of a satellite in an image is shifted relative to its geometric center, the center of mass, due the solar phase angle. According to \cite{Lindegren1977} a spherical object with a light scattering in its surface causes an offset in its positions according to Equations (\ref{eq:fase1}) where $i$ is the solar phase angle, $r$ is the apparent radius of the satellite, $Q$ is the position angle of the sub-solar point in the tangential plane and $C(i)$ is a parameter related to the reflectance model adopted.      

\begin{equation}
\left(\bfrac{\Delta\alpha\cos\delta}{\Delta\delta}\right) = C(i)r\sin(i/2)\left(\bfrac{\sin Q}{\cos Q}\right) \label{eq:fase1}
\end{equation}

Here we adopted the Lambertian sphere modelling \citep{Lindegren1977}. But since the satellites' radii may be different, and considering that the direction of the Sun and of the relative motion may not coincide,  the relative distances of the photo and geometric centres may vary differently, so that the associated central instants may be different too, causing an offsset in the observed central instant which is associated to the photo and not to the geometric center. For example, in the case of the Galilean moons, for an approximation with a phase angle of 10 degrees the central instant can be shifted by up to 6 seconds, a significant value.      

The atmospheric refraction causes a shift in the target position towards the zenith direction that increases at higher zenital distances. However, once both satellites have almost the same zenital distance and very close positions in an approximation it has a small, but non-negligible effect in the distances and in the central instant (less than 2 seconds for $z \approx 70^{o}$). More details about its implementation can be found in \cite{Stone1996}.   

The aberration causes a position shift toward the direction of the instantaneous velocity vector of the observer \citep{Green}. Due to the small apparent distance between satellites, the effect is very small in the central instant, less than 0.0005 seconds for diurnal aberration and 0.0145 seconds for annual aberration.

\section{Mutual approximations of Galilean satellites: results of the campaign of 2014-2015} \label{results}

\subsection{Programme}

The accurate prediction of mutual phenomena depends on how precise the ephemeris is and upon our knowledge of the size of the satellites. In the absence of one of these conditions, we may miss an event foreseen to be visible at a certain location. None of these conditions matter in our case. One can easily predict mutual approximations, even using poor ephemeris with precision in the arcsecond level only. By setting a threshold value of at least some arcseconds, we avoid selecting possible mutual occultations or situations where the apparent distances between the satellites are too small for centroid measurement purposes.

\begin{figure}
\includegraphics[height=07cm]{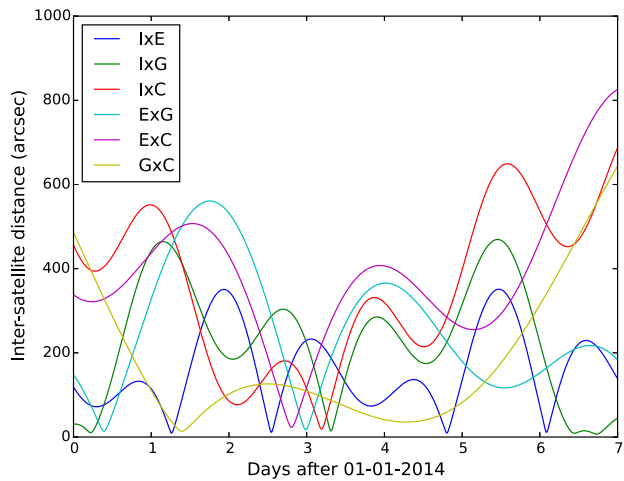}
\caption{Inter-satellite apparent sky distances of the Galilean satellites over a week. I stands for Io, E for Europe, G for Ganymede and C for Callisto. Every minima in this graphic is a mutual approximation between two satellites. We used the NOE-5-2010-GAL ephemeris obtained in its topocentric form for OPD.}
\label{Fig_002}
\end{figure}

The observational campaign for the mutual approximations between the Galilean satellites was carried out in Brazil between 2014 and 2015. The predictions for these events were made with the topocentric ephemeris for the OPD observatory using the NAIF SPICE\footnote{Website: \url{http://naif.jpl.nasa.gov/naif/}} platform, ephemeris NOE-5-2010-GAL, derived from \cite{Lainey2009}, and DE430. Figure \ref{Fig_002} plots the inter-satellite apparent distances, in arcseconds, over a week for the six possibles combinations between these satellites. Every minima is a mutual approximation. However, in order to not pick up a prohibitive number of events, we only selected the approximations for which the impact parameter was smaller than 30 arcseconds, improving even further the precision premium.

We selected all the visible events for the OPD observatory with elevation above 20 degrees and with a distance to the Jupiter's limb greater than 10 arcseconds. The predictions were spread in 58 nights, with 65 approximations, selected in fifteen months between 2014 and 2015. We attempted observations for all these events. From these, only 36 events could be observed with success, the others were lost due to bad weather conditions.

\subsection{Observations}

Our observations were made at Observat\'orio do Pico dos Dias (OPD, IAU code 874)\footnote{Website: \url{http://www.lna.br/opd/opd.html}} located at geographical longitude $-45^{o}~34'~57''$, latitude $-22^{o}~32'~04''$ and an altitude of $1864~m$. The telescope used was the $0.6~m$ diameter Zeiss telescope. It is a manual pointing Cassegrain telescope with focal ratio f/12.5.

For all the observations in this work the time, in UTC, was calibrated by a GPS and recorded in the FITS image's header. For observations which the GPS is not an option, a time's calibrator software can be used such as \emph{Dimension 4} \footnote{Website: \url{http://www.thinkman.com/dimension4/}}. Tests comparing this software with GPS indicate that the time precision is on the order of 20 $ms$.

The CCD camera utilized in all observations was the Andor Ikon-L with 2048 x 2048 square pixels of 13.5 $\mu m$. This camera added to the Zeiss telescope has a field of view of 12.63' x 12.63'. The filter chosen was the narrow-band filter centered at 889 $nm$ (region of methane absorption), with a width of 15 $nm$. In this specific wavelength, the methane in Jupiter's upper atmosphere strongly absorb the light causing the planet's albedo to drop to 0.1 in this spectral region as pointed out by \cite{Karkoschka1994,Karkoschka1998}.

\begin{figure}
\includegraphics[height=05.4cm]{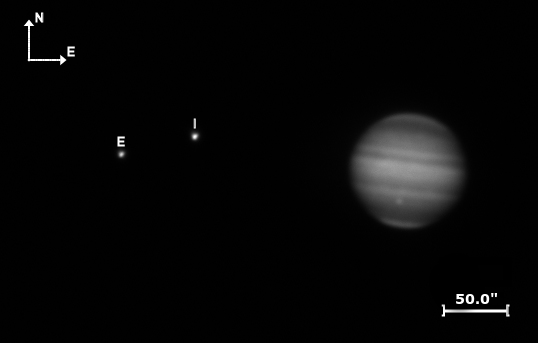}
\caption{Image of Jupiter, Io and Europa obtained with the 0.6 m diameter Zeiss telescope, equipped with a methane filter in February $3^{th}$ of 2014. The planet and the satellites show about the same brightness due to the use of the narrow-band filter, centred at $\lambda = 889~nm$ with $15~nm$ width.}
\label{Fig_001}
\end{figure}

Although the observations in this wavelength are very efficient to eliminate the scattered light from Jupiter, the albedo of the Galilean satellites did not change much \citep{Karkoschka1994}. Because of this, the brightness of Jupiter is nearly the same of that of the satellites in this wavelength as can be seen in the Figure \ref{Fig_001}.

We summarized the specifications of the telescope, camera and filter utilized in these observations in Table \ref{tb_001}.  

\begin{table}
\caption{Specifications of the telescope, CCD camera and filter.}
\begin{tabular}{cc}
\hline
Diameter of primary mirror& $0.60~m$\\
Focal ratio& $f/12.5$\\
CCD FOV& $12.63'$ x $12.63'$\\
Pixel size& $13.5$ x $13.5~\mu m^{2}$\\
Size of CCD array&$2048$ x $2048$ \\
Pixel-scale&$0\farcs34$ or $0\farcs37~pixel^{-1}$ \\
Filter&$889~nm~(\Delta = 15~nm)$\\
\hline
\label{tb_001}
\end{tabular}
\end{table}

Table \ref{tb_002} contains the observational characteristics for the 36 nights analysed in this work, where 14 are mutual approximations (See Section \ref{result}) and 22 are mutual approximation observed from mutual occultations data of the 2014-2015 mutual phenomena campaign (See Section \ref{5.2}). We list the mutual approximation targeted, the seeing of the night, the zenithal distance ($z_0$) for the central instant, the solar phase angle ($i$) and the sub-solar point in the tangential plane ($Q$). The solar phase angles and the sub-solar point in the tangential plane are the same for both satellites and do not change during the approximation, which lasts typically less than two hours. The observational information about the 5 nights in the 2009 mutual phenomena campaign utilized in this work can be found in \cite{Diasoliveira2013} (See Section \ref{5.1}).

\begin{table}

\caption{Approximations and observation conditions}
\begin{tabular}{cccccc}
\hline
Date &Event &Seeing &$z_0$ & $i$ & $Q$ \\
(d-m-y) & &(arcsec) &($^o$) & ($^o$) & ($^o$) \\
\hline
03-02-2014 & IaE & 1.9 & 56.74 &  5.80 & 275.44 \\
05-02-2014 & EaG & 2.1 & 49.58 &  6.27 & 275.36 \\
07-02-2014 & IaE & 1.8 & 51.87 &  6.61 & 275.28 \\
19-02-2014 & IaG & 2.8 & 45.76 &  8.39 & 274.84 \\
27-02-2014 & IaE & 1.9 & 47.23 &  9.32 & 274.71 \\
18-03-2014 & IaE & 2.4 & 46.68 & 10.73 & 274.63 \\
07-04-2014 & IaE & 2.4 & 51.73 & 10.99 & 275.15 \\
20-04-2014 & EaG & 2.3 & 52.12 & 10.58 & 275.75 \\
21-04-2014 & GaC & 2.2 & 51.38 & 10.54 & 275.77 \\
21-04-2014 & IaG & 2.2 & 64.52 & 10.54 & 275.77 \\
15-10-2014 & GaC & 1.3 & 66.32 &  9.70 & 108.26 \\
15-10-2014 & IaE & 1.3 & 66.32 &  9.70 & 108.26 \\
29-10-2014 & IaG & 1.5 & 54.46 & 10.45 & 108.67 \\
02-11-2014 & IaC & 2.0 & 68.93 & 10.60 & 108.67 \\
19-11-2014 & EaC & 1.3 & 41.34 & 10.69 & 108.86 \\
02-02-2015 & GaC & 1.7 & 42.82 &  1.01 &  98.03 \\
22-02-2015 & IaE & 1.7 & 39.58 &  3.17 & 290.47 \\
24-02-2015 & IaG & 1.8 & 40.01 &  3.59 & 289.99 \\
26-02-2015 & IaE & 1.6 & 64.54 &  4.14 & 289.47 \\
27-02-2015 & GaC & 1.6 & 40.35 &  4.16 & 289.46 \\
27-02-2015 & IaG & 1.6 & 40.60 &  4.17 & 289.45 \\
03-03-2015 & IaG & 1.6 & 56.01 &  4.95 & 288.90 \\
24-03-2015 & GaC & 2.2 & 41.53 &  8.19 & 287.38 \\
25-03-2015 & IaE & 1.6 & 40.65 &  8.46 & 287.27 \\
02-04-2015 & IaE & 1.5 & 52.27 &  9.14 & 287.07 \\
11-04-2015 & EaG & 2.1 & 41.50 & 10.07 & 286.82 \\
13-04-2015 & IaE & 1.5 & 43.80 & 10.21 & 286.79 \\
17-04-2015 & IaC & 1.4 & 45.35 & 10.42 & 286.72 \\
18-04-2015 & GaC & 1.4 & 61.45 & 10.42 & 286.72 \\
18-04-2015 & IaG & 1.5 & 45.31 & 10.45 & 286.73 \\
19-04-2015 & EaG & 1.4 & 59.89 & 10.46 & 286.73 \\
21-04-2015 & IaE & 1.4 & 68.34 & 10.57 & 286.70 \\
25-04-2015 & IaG & 2.7 & 48.86 & 10.41 & 286.70 \\
26-04-2015 & IaE & 2.0 & 40.67 & 10.74 & 286.69 \\
29-04-2015 & IaG & 1.9 & 57.72 & 10.81 & 286.70 \\
03-05-2015 & IaE & 2.0 & 52.39 & 10.84 & 286.71 \\
\hline
\label{tb_002}
\end{tabular}
{\it Note.} For each event, we have the day, month and year, the satellites designated by their initials (capital letters), where 'a' stands for approximation. We also give the seeing, the zenithal distance ($z_0$), the solar phase angle ($i$) and the position angle of the sub-solar point in the tangential plane ($Q$).    
\end{table}

\subsection{Photocentre measurement}

\begin{figure}
\includegraphics[height=07cm]{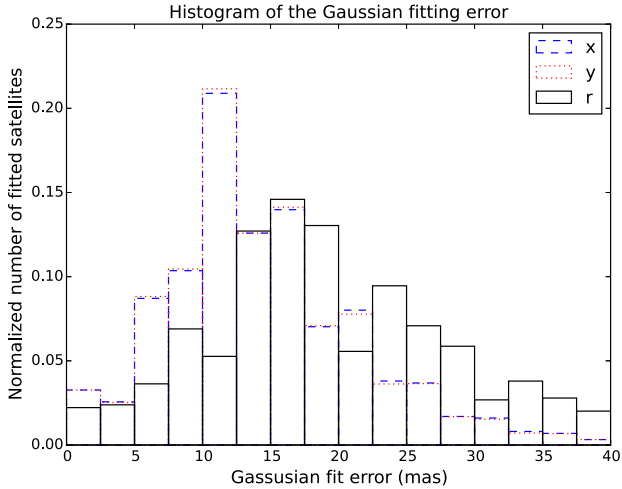}
\caption{Normalized histogram for the centroid determination error, in mas, for $x$, $y$ and $ r = \sqrt{x^2+y^2}$, from all measured images.}
\label{Fig:hist_PRAIA}
\end{figure}

Firstly, all the images were corrected by bias and flat-field by means of standard procedures using the IRAF\footnote{Website: \url{http://iraf.noao.edu/}} package \citep{BeS1981}. The centroid of the satellites were determined using the PRAIA package described in \cite{Assafin2011}.
The PRAIA package measures the satellite coordinates (x, y)  in the image with a two-dimensional circular symmetric Gaussian fit within a radius of one Full Width Half Maximum (FWHM = seeing). The typical error in the centroid measurement for our images was 12 mas. Figure \ref{Fig:hist_PRAIA} is the normalized histogram for the centroid determination error, in mas, for $x$, $y$ and $r = \sqrt{x^2+y^2}$, from all measured images.

\subsection{Results} \label{result}

We used the model described in Section \ref{model} to determine the central instant $t_0$ of the observed mutual approximations. The current sky path of Jupiter is not crowded of stars. Also, Jupiter’s brightness made us use a short exposure time and a narrow-band filter. Because of that there was not enough number of reference stars in the images for an usual CCD astrometry. Even so, for evaluation purposes, we also determined the impact parameter $d_0$  and the relative velocity $v_r$ following the procedures given in Sect. \ref{Anexo_A} with the help of an ephemeris.

We separate our results in two different groups. Group 1 contains our best results. It consists of observations made in good sky conditions with no gaps along the event. Group 2 has gaps in the distance curve, which may present more noise than in Group 1 due to poor atmosphere conditions. Some observations in the later group were virtually made during the 2014-2015 Jupiter equinox. Although they were not mutual occultations, the satellites approached each other by very small distances, about 2 arcseconds.

\subsubsection{Mutual approximations: Group 1} \label{Group_1}

This campaign started in the beginning of 2014. Since the mutual phenomena started in the middle of the same year, only a few mutual approximations were observed without gaps in the center of the curves due to the apparent proximity of the satellites. The group 1 is composed of 8 events. These approximations are similar to the ones that will be observed after the mutual phenomena campaign, this makes them the focus of this paper.  

\begin{figure}
\includegraphics[height=07cm]{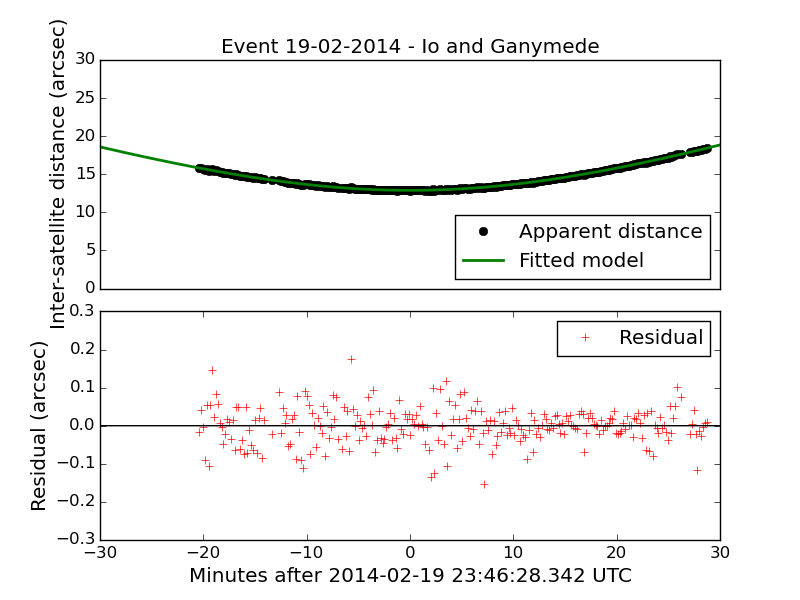}
\caption{Apparent sky distances between Io and Ganymede in the mutual approximation of February, 19th 2014. Inter-satellite distances, $d$, are in black and model fitted ones in green. The exposure time utilized was 3 seconds. Notice that the fit is actually done in $d^2$, using Eq. (\ref{eq:Ad4}).}
\label{Fig_result_1}
\end{figure}

As an example we display the approximation between Io and Ganymede that occurred in February, 19th of 2014. The comparison between the fitted (green) and observed (black) apparent distances is shown in Figure \ref{Fig_result_1}. The residual dispersion is also illustrated in the bottom (red crosses), in the sense \emph{"fitted minus observed"}.

We show in Table \ref{tb:result_1} the central instant,  in hours, minutes and seconds in UTC and the error in seconds. We also list the error in milliarcseconds by the use of the ephemeris relative velocity in milliarcseconds per second. These values are listed in column $E_{t0}$. We compared the results with the ephemeris published by the IMCCE, currently considered the most accurate representative for the Jovian system, in the sense \emph{"observations minus ephemeris"}. 

The average precision obtained for the central instant is 0.56 seconds or 3.42 mas (using the relative velocity in each event obtained with the ephemeris).

\begin{table}
\caption{The central instant for the Group 1 mutual approximations and comparison with the ephemeris (see text in Sect. \ref{Group_1}).}
\begin{center}
\begin{tabular}{ccccrr}
\hline
Date &Event &$t_{0}$  & $E_{t0}$ &\multicolumn{2}{c}{$\Delta t_{0}$} \\
(d-m-y) & & (hh:mm:ss) & (mas) & (s) & (mas)  \\
\hline
03-02-14	&	 IaE	&	    03:18:47.42 (0.20)	& 1.55	& +4.37	&	+33.92	\\
05-02-14	&	 EaG	&	    23:27:50.65 (0.66)	& 4.04	& +2.86	&	+17.53	\\
19-02-14	&	 IaG	&	    23:46:35.40 (0.76)	& 5.78	& +2.31	&	+17.59	\\
27-02-14	&	 IaE	&	    22:34:27.89 (0.10)	& 0.79	& --1.16&	--9.16	\\
07-04-14	&	 IaE	&	    22:35:27.89 (0.19)	& 1.46	& --0.35&	--2.68	\\
20-04-14	&	 EaG	&	    21:47:40.57 (0.52)	& 1.85	& --9.70&	--34.47	\\
21-04-14	&	 GaC	&	    21:41:53.56 (1.01)	& 4.74	& +1.71	&	+8.04	\\
21-04-14	&	 IaG	&	    23:13:56.69 (1.01)	& 5.45	& --2.73&	--14.74	\\

\hline
\label{tb:result_1}
\end{tabular}
\end{center}
{\it Note.} $t_0$ is the central instant of the mutual approximations. Column $E_{t0}$ lists the central instant error in mas by the use of the ephemeris relative velocities in mas per second. For each event, we have the day, month and year, the satellites designated by their initials (capital letters), where 'a' stands for approximation. The column $\Delta t_0$ is the comparison between the observations and the ephemeris, here the  $NOE-5-2010-GAL$ from IMCCE plus DE430, derived from \cite{Lainey2009}, in the sense \emph{"observations minus ephemeris"} in seconds and in mas by the use of the relative velocities in each event.  
\end{table}

\subsubsection{Mutual approximations: Group 2} \label{Group_2}

A mutual approximation can last a couple of hours. This makes it possible to acquire many observations and fit the model even without the full coverage of the approximation. Logically, these results will not be as good as if we could observe all the event. We have 6 mutual approximations in this group. A natural cause for the random observational gaps in the curves were bad weather or very bad seeing.   

\begin{figure}
\includegraphics[height=07cm]{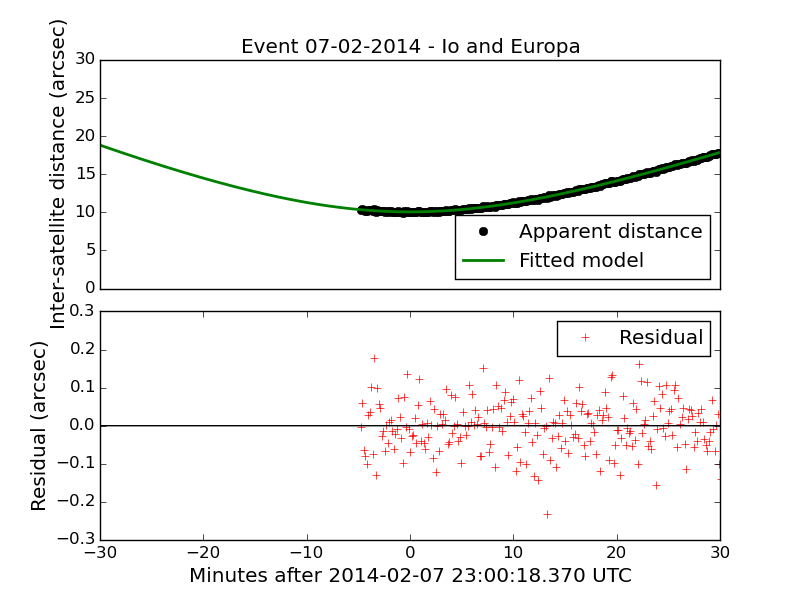}
\caption{Apparent sky distances between Io and Europa in the mutual approximation of February, 07th 2014. Inter-satellite distances, $d$, are in black and model fitted ones in green. The exposure time utilized was 3 seconds. Notice that the fit is actually done in $d^2$, using Eq. (\ref{eq:Ad4}).}
\label{Fig_result_2}
\end{figure}

Two examples can be displayed. One is when the weather prevented the observation of part of the event such as in the approximation between Io and Europa that occurred in February, 07th of 2014. In Figure \ref{Fig_result_2} we show the comparison between the fitted and observed apparent distances. Notice the absence of positions in the left side of the event.

\begin{figure}
\includegraphics[height=07cm]{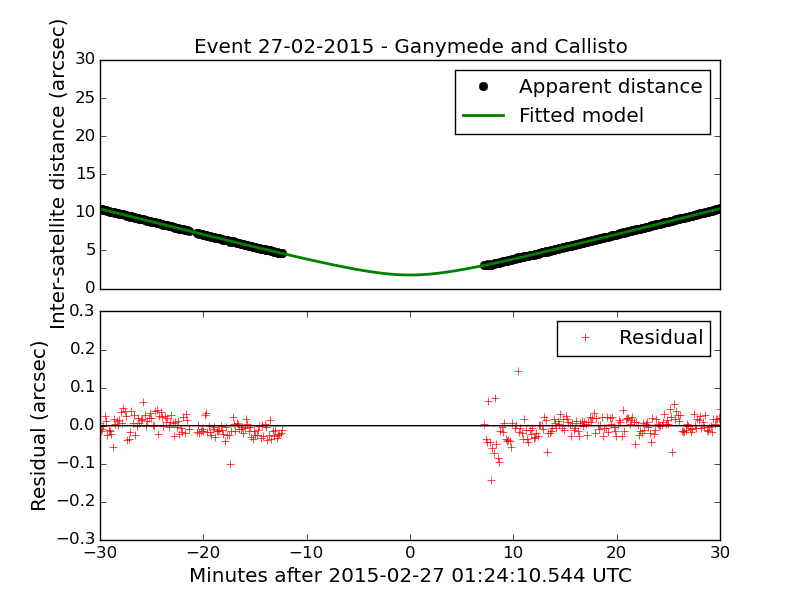}
\caption{Apparent sky distances between Ganymede and Callisto in the mutual approximation of February, 27th 2015. Inter-satellite distances, $d$, are in black and model fitted ones in green. The exposure time utilized was 3 seconds. Notice that the fit is actually done in $d^2$, using Eq. (\ref{eq:Ad4}).}
\label{Fig_result_3}
\end{figure}

The second example is when the approximation occurred during the mutual phenomena campaign at the planet's equinox. Near the central instant of the approximation, the satellites are so close together that it is impossible to obtain a centroid for each satellite in the images. An example is the approximation between Ganymede and Callisto that occurred in February, 27th of 2015. In Figure \ref{Fig_result_3} we show the comparison between the fitted and observed apparent distances. Notice the absence of positions near the central instant.

We show in Table \ref{tb:result_2} the central instant for these six approximations, in the same manner as in Table \ref{tb:result_1} for the Group 1. We also compared the parameters with the IMCCE ephemeris.   

The average precision obtained for the central instant is 2.02 seconds or 14.15 mas (using the relative velocity in each event obtained with the ephemeris). Notice that even here the precisions are still much better than those obtained with usual CCD astrometry.

\begin{table}
\caption{Central instant for the Group 2 mutual approximations and comparison with the ephemeris (see text in Sect. \ref{Group_2})}
\begin{center}
\begin{tabular}{ccccrr}
\hline
Date &Event &$t_{0}$  & $E_{t0}$ &\multicolumn{2}{c}{$\Delta t_{0}$} \\
(d-m-y) & & (hh:mm:ss) & (mas) & (s) & (mas)  \\
\hline
07-02-14 &  IaE & 23:00:17.29 (4.94)  &   40.97   & --1.38  &   --11.46 \\
18-03-14 &  IaE & 22:43:23.26 (2.52)  &   16.06   & --7.93  &   --50.54 \\
27-02-15 &  GaC & 01:24:10.29 (1.42)  &   8.14    & +0.71   &     +4.07 \\
11-04-15 &  EaG & 22:07:13.35 (1.13)  &   5.76    & --2.60  &   --13.25 \\
13-04-15 &  IaE & 23:44:40.98 (1.12)  &   8.60    & --1.78  &   --13.67 \\
19-04-15 &  EaG & 01:17:48.89 (1.00)  &   5.15    & --4.10  &   --21.12 \\
\hline
\label{tb:result_2}
\end{tabular}
\\{\it Note.} Same as in Table \ref{tb:result_1}.  
\end{center}
\end{table}

How does the gaps in these distance curves affect the determination of the central instant? We addressed this question by simulating gaps in the Group 1 approximations reported in the previous section. As example, we illustrate the simulation using the event between Io and Ganymede in February 19th 2014. Similar conclusions are drawn for the other simulations. The event took 1 hour, thus a 6-minutes gap represents 10\% of the entire curve.  

\begin{figure}
\includegraphics[height=07cm]{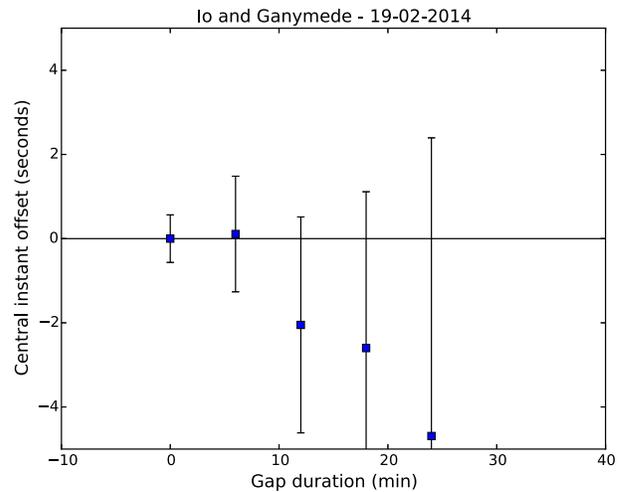}
\caption{Simulation of gaps in the beginning of the Group 1 mutual approximation between Io and Ganymede of February, 19th 2014. Offsets are in the sense \emph{"with gap minus without gap"}.}
\label{Fig:lado}
\end{figure}

\begin{figure}
\includegraphics[height=07cm]{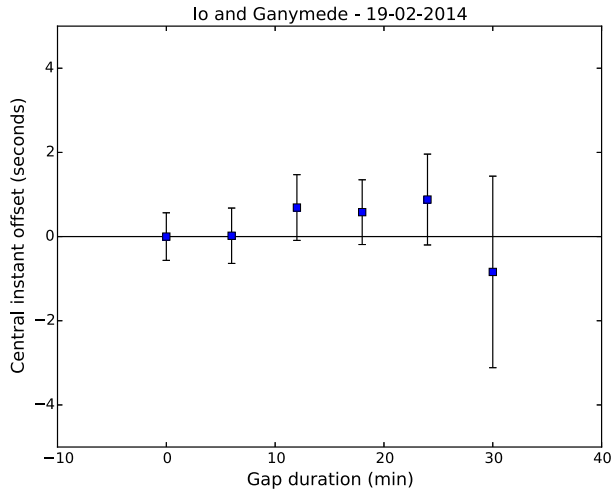}
\caption{Simulation of gaps in the central part of the Group 1 mutual approximation between Io and Ganymede of February, 19th 2014. Offsets are in the sense \emph{"with gap minus without gap"}.}
\label{Fig:central}
\end{figure}

We explore two scenarios. (1) Removing points from the beginning, or end, of the curve. This affects the central instant error, even though the central instant value is close (within the errors) to the one obtained with the complete curve (Figure \ref{Fig:lado}); (2) Removing points in the central part of the curve. The central instant  precision and value are nearly unaffected (Figure \ref{Fig:central}). The offsets are consistent within the errors.

These simulations confirm the slight deterioration observed in the errors of the parameters of the incomplete mutual approximations of Group 2, as compared to those from the Group 1. But the simulations also indicate that the obtained central instant are practically unaffected with regard to the ones that would be derived with complete curves, within the errors. Therefore, in principle even incomplete curves of mutual approximations should not be discarded.

\subsubsection{Results for the impact parameter and relative velocity} \label{A3}

In Table \ref{tb:secondary_fit} we list the impact parameters and the relative velocity of our 14 mutual approximations. They are listed in milliarcseconds, and milliarcseconds per second, respectively. We also list the comparison of these parameters with the ephemeris in the sense \emph{"observations minus ephemeris"}. Both observed and ephemeris results were obtained by fitting the model of Section \ref{A2} to the respective observed and ephemeris apparent distances. The pixel scale determined is also listed. The nominal pixel scale of the instrumental set was 0.34 or 0.37 mas per pixel depending of the instrumental configuration of the night. The last column is an identification flag where 1 stands for "Group 1  mutual approximations" and 2 stands for "Group 2 mutual approximations". Notice that, in the near future, a new reduction can be made using more precise ephemeris. This will allow the confirmation of ours results.

\begin{table*}
\caption{Fitted impact parameter and relative velocity and comparison between observations and the ephemeris}
\begin{tabular}{cccccccc}
\hline
Date &Event & $d_0$ & $v_r$ & $\Delta d_0$ & $\Delta v_r$ & $ P_s$ & Id.\\
(d-m-y) & & (mas) & (mas/s) & (mas) & (mas/s) & ("/px) &\\
\hline
03-02-2014 	&	 IaE 	&	  9908.66 (07.93) 	&	 7.81 (0.31) 	&	--17.18	&	+0.05	&	 0.37214 (0.00011) 	&	 1\\
05-02-2014 	&	 EaG 	&	 15014.25 (09.61) 	&	 6.24 (0.35) 	&	+51.00	&	+0.04	&	 0.37478 (0.00005) 	&	 1\\
07-02-2014 	&	 IaE 	&	 10080.02 (12.18) 	&	 8.36 (0.71) 	&	--73.26	&	+0.02	&	 0.37154 (0.00028) 	&	 2\\
19-02-2014 	&	 IaG 	&	 12995.06 (06.08) 	&	 7.55 (0.22) 	&	+0.43	&	+0.07	&	 0.37288 (0.00087) 	&	 1\\
27-02-2014 	&	 IaE 	&	  9809.83 (13.64) 	&	 7.96 (0.38) 	&	--7.61	&	--0.03	&	 0.34943 (0.00007) 	&	 1\\  
18-03-2014 	&	 IaE 	&	  7984.82 (29.31) 	&	 6.52 (0.71) 	&	+7.69	&	+0.09	&	 0.34971 (0.00014) 	&	 2\\
07-04-2014 	&	 IaE 	&	  8866.39 (11.83) 	&	 7.74 (0.34) 	&	--24.28	&	+0.04	&	 0.36115 (0.00012) 	&	 1\\
20-04-2014 	&	 EaG 	&	  8425.61 (03.46) 	&	 3.56 (0.08) 	&	+4.27	&	+0.02	&	 0.36161 (0.00033) 	&	 1\\
21-04-2014 	&	 GaC 	&	 18052.70 (02.97) 	&	 4.72 (0.16) 	&	--13.71	&	+0.03	&	 0.36233 (0.00057) 	&	 1\\
21-04-2014 	&	 IaG 	&	  9250.29 (07.25) 	&	 5.48 (0.52) 	&	--24.25	&	+0.09	&	 0.36235 (0.00064) 	&	 1\\
27-02-2015 	&	 GaC 	&	 1763.02 (25.99)  	&	 5.76 (0.19) 	&	--25.97	&	+0.03	&	  0.34873 (0.00005) 	&	 2\\
11-04-2015 	&	 EaG 	&	 2173.79 (19.42)  	&	 5.16 (0.36) 	&	--22.31	&	+0.09	&	  0.34837 (0.00017) 	&	 2\\
13-04-2015 	&	 IaE 	&	 1358.27 (42.35)  	&	 7.73 (0.25) 	&	--61.96	&	+0.05	&	  0.34860 (0.00011)  	&	 2\\
19-04-2015 	&	 EaG 	&	 2126.17 (31.88)  	&	 5.19 (0.37) 	&	--75.47	&	+0.07	&	  0.34818 (0.00027) 	&	 2\\

\hline
\label{tb:secondary_fit}
\end{tabular}
\\{\it Note.} The impact parameter, $d_0$ and the relative velocity $v_r$ of the mutual approximations. The columns $\Delta$ are the comparisons between the observations and the ephemeris parameters, here the  $NOE-5-2010-GAL$ from IMCCE plus DE430, derived from \cite{Lainey2009}, in the sense \emph{"observations minus ephemeris"}. For each event, we have the day, month and year, the satellites designated by their initials (capital letters), where 'a' stands for approximation. The Pixel-scale, $P_s$, determined with the ephemeris is, also, listed. In the Id. column 1 stands for "Group 1 mutual approximations" and 2 stand for " Group 2 mutual approximations".  
\end{table*}

\section{Mutual occultations inside mutual approximations} \label{comparar}

Mutual approximations and mutual phenomena - occultations in particular - share the same concepts of orbital geometry, though based in very distinct measuring techniques, with the last being a consolidated and most precise method for measuring distances between natural satellites. It would be very interesting if we could compare the performance of both methods in an equal basis. Indeed this is possible because, in a broader sense, any mutual occultation is always contained in a mutual approximation. The only drawback is that the same set of useful observations to be fitted in mutual occultations, when the satellites are too close together, is exactly the set that must be discarded in the mutual approximations, and vice-versa. Even so, this still makes an interesting comparison, because the instrumental and astronomical observational conditions are quite the same, and the independence of the observational sets has a relevance of its own.

\subsection{A comparison from observations of Jupiter's 2009 equinox}\label{5.1}

For this comparison, we used the data of the mutual phenomena campaign of 2009 of \cite{Diasoliveira2013}. We utilized images acquired before and after five occultations originally designed for albedo determination. \cite{Arlot2014} also determined the geometric parameters of these five occultations using the same light curves as \cite{Diasoliveira2013}, however obtaining slightly different results. Here we compare the results derived from the mutual approximations with the results of \cite{Arlot2014} and the NOE-5-2010-GAL ephemeris. The comparison is displayed in Table \ref{tb:compar_phemu}. As shown in Table \ref{tb:compar_phemu}, the results for the mutual approximation method agrees with the results of \cite{Arlot2014} and the NOE-5-2010-GAL, within the errors. This highlights the strength of mutual approximations.

\begin{table*}
\caption{Comparison between the central instant for five mutual approximations and occultations observed in the 2009 equinox of Jupiter}
\begin{tabular}{cccccccccc}
\hline
Date & Event & \multicolumn{2}{c}{[1] error} &\multicolumn{6}{c}{Central instant difference} \\
         & &   &      &\multicolumn{2}{c}{[1] - [2]} & \multicolumn{2}{c}{[1] - [3]} & \multicolumn{2}{c}{[2] - [3]} \\
 (d-m-y) & & (s) & (mas) & (s) & (mas) & (s) & (mas) & (s) & (mas) \\
\hline
 09-05-09 & IaE & 3.23 &23.84 & --0.50 & --3.65 &  +0.40 &  +2.98 &  +0.90   &	   +6.63  \\
 28-05-09 & IaE & 3.62 &22.55 & +2.28  & +14.16 & --2.31 &--14.34 & --4.59   &	 --20.49  \\
 22-06-09 & IaE & 4.72 &26.57 & --2.25 & --12.77& --4.51 &--25.59 & --2.26   &	 --12.82  \\
 06-07-09 & IaE & 3.02 &16.01 & --2.98 & --15.61& --4.66 &--24.35 & --1.67   &	  --8.74  \\
 07-08-09 & IaE & 3.59 &13.46 & +1.96  &  +7.33 & --7.07 &--26.40 & --9.03   &	  -33.72  \\
\hline
\end{tabular}
\label{tb:compar_phemu}
\\{\it Note.} [1] Mutual approximation; [2] \cite{Arlot2014}; [3] Ephemeris, NOE-5-2010-GAL from IMCCE plus DE430. The average error of the approximations are 3.63 seconds (20.47 mas) for the central instant. Notice that the difference between the approximations and the mutual phenomena of \cite{Arlot2014} is smaller than the errors of the central instant of the mutual approximation, and has the same order that the difference between \cite{Arlot2014} and the ephemeris.
\end{table*}

\subsection{Mutual approximations in the observations of Jupiter's 2014-2015 equinox}\label{5.2}

The results for the mutual phenomena campaign of 2014-2015 are still being processed. However we list here the central instant for the mutual approximations derived from the observations before and after the mutual phenomena themselves. These results can be used, in the near future, for comparison with the results from the mutual occultations of this campaign. It is important to stress that, for this scenario, the precision of the mutual approximation results is below its capacity, once we will always have an absence of points around the central instant. 

These results are shown in Table \ref{tb:phemu_2}, where we list the central instant and its error obtained from our analysis for 22 mutual approximations and the comparison with the ephemeris, similar as in Table \ref{tb:result_1}

\begin{table*}
\caption{Results for the mutual approximations for the 22 occultations observed in 2014-2015 and comparison with the ephemeris.}
\begin{center}
\begin{tabular}{ccccrr}
\hline
Date &Event &$t_{0}$  & $E_{t0}$ &\multicolumn{2}{c}{$\Delta t_{0}$} \\
(d-m-y) & & (hh:mm:ss) & (mas) & (s) & (mas)  \\
\hline
15-10-14 &GaC & 	07:07:07.26 (3.55)  & +4.80   & 14.60	& +19.75  \\
15-10-14 &IaE & 	07:07:54.19 (2.48)  & +0.02   & 17.54	& +0.16   \\
29-10-14 &IaG & 	07:07:24.26 (4.34)  & --2.05  & 22.17	& --10.45 \\
02-11-14 &IaC & 	06:02:34.11 (9.19)  & --2.82  & 16.75	& --5.14  \\
19-11-14 &EaC & 	07:37:52.48 (3.61)  & +11.79  & 18.32	& +59.86  \\
02-02-15 &GaC & 	02:24:31.55 (2.57)  & --3.56  & 14.99	& --20.74 \\
22-02-15 &IaE & 	02:07:53.57 (1.32)  & +6.24   & 7.21	& +34.07  \\
24-02-15 &IaG & 	01:44:46.97 (6.43)  & +2.72   & 53.15	& +22.48  \\
26-02-15 &IaE & 	22:21:26.26 (2.53)  & +1.32   & 23.72	& +12.40  \\
27-02-15 &IaG & 	02:20:23.81 (1.30)  & --1.63  & 8.54	& --10.68 \\
03-03-15 &IaG & 	04:08:13.38 (3.85)  & --1.46  & 32.28	& --12.24 \\
24-03-15 &GaC & 	00:14:15.44 (5.08)  & --0.02  & 27.19	& --0.09  \\
25-03-15 &IaE & 	23:34:49.47 (10.46) & --9.13  & 65.38	& --57.08 \\
02-04-15 &IaE & 	01:43:57.43 (1.86)  & +4.26   & 11.83	& +27.09  \\
17-04-15 &IaC & 	23:47:03.29 (1.26)  & +3.00   & 6.28	& +14.92  \\
18-04-15 &GaC & 	01:32:21.21 (1.16)  & --0.58  & 5.80	& --2.89  \\
18-04-15 &IaG & 	20:54:40.80 (2.06)  & +0.07   & 11.39	& +0.39   \\
21-04-15 &IaE & 	01:55:02.68 (1.16)  & --1.61  & 8.56	& --11.86 \\
25-04-15 &IaG & 	23:45:24.78 (1.51)  & +1.23   & 8.99	& +7.30   \\
26-04-15 &IaE & 	21:24:57.44 (4.12)  & +0.12   & 27.29	& +0.79   \\
29-04-15 &IaG & 	00:28:55.49 (4.36)  & --14.96 & 29.87	& --102.50\\
03-05-15 &IaG & 	23:39:20.87 (1.53)  & +2.35   & 10.20	& +15.70  \\
\hline
\label{tb:phemu_2}
\end{tabular}
\\{\it Note.} Same as in Table \ref{tb:result_1}.  
\end{center}
\end{table*}

\section{Conclusion} \label{conclusao}

In this paper we presented a method to measure the central instant in an approximation  between natural satellite pairs. Instead of being restricted to the particular configuration of mutual occultations - which only occur during the equinox of the central planet - in mutual approximations (as we call the method)  we can make observations every time when the satellites don't cross each other in the sky, but rather approach each other up to a minimum apparent distance, which is at least larger than the sum of their radii, or in practice larger than the seeing - a special geometry that recurrently occurs for natural satellites. In this method the relative motion of the satellites is essentially described by the same geometric parameters as in a mutual occultation -- central instant, impact parameter and relative velocity. But only the central instant can be truly determined independently of any reference system, and so we consider it the main result of the method. Here astrometry is the technique used to directly measure typically very short distances (less than 85") with very small errors due to the precision premium, while differential photometry is the technique used in mutual phenomena.

We successfully applied the method to the Galilean moons using CCD observations made in 2014 and 2015. We compared ours results with the ephemeris NOE-5-2010-GAL from IMCCE. Using old observations from the 2009 equinox of Jupiter, we also compared the performance of mutual approximations with published mutual phenomena results from that campaign.

The frequency of these approximations depends only of the orbital period of the satellites. In the case of the Galilean moons, a couple of days or so. Because there is no need of reference stars for the astrometry even small telescopes can be used. Because the events may last for hours there is no need of a high cadence in time between the images, even tens of seconds would be ok.

The mutual approximations extend the possibility of obtaining relative distances with precision of a few mas, to periods where there are no occultations or eclipses. This means getting a relative position with an error about 10 mas for every observed event along the visibility period. In the case of the Galilean moons we obtained a precision of 0.56 seconds for the central instant when the whole approximation curve of distances was observed and a precision of 1.52 seconds for the central instant when there were gaps along the curve or around the central instant.

The  high  precision results obtained in this work for the Galilean moons benefited from:  (i)  the precision premium from very small field astrometry; (ii) from the use of a narrow band filter centred in a methane absorption region, eliminating the scattered light of Jupiter (this filter was also used in the reported mutual phenomena observations); (iii) the use of an adequate telescope/detector/exposure configuration set, allowing for imaging the satellites with high S/N (signal/noise) ratios, but avoiding saturation. 

The instrumental distortions due to the non flatness and non parallelness of the filter and the CCD cover glass, and their distance from the CCD chip affect the global reduction of the entire FoV of the CCD to some extent. However, due to the very small distance between both satellites (smaller than 30 arc seconds) these distortions can be neglected here.

The error in the measurement of the centroids, and thus of the distances, due to the effects of low/high albedo regions in the surface of the satellites, is presently unknown. However, taking the surface illumination resulting from the solar phase angle geometry as an extreme example, and using the relations in Section \ref{reduc}, we find that we need nearly a 5 degrees phase angle to change the photocentre by 10 mas. This corresponds to a zero albedo circular region of 100 km radius. Indeed, craters or volcanoes in many of the Galilean moons are features of this size \citep{Faure2007}. However, they represent a variation of only 0.1 in the surrounding surface albedo. So, in principle, at least for the Galilean moons, this effect can be neglected. In the special case of Io, the volcanoes can affect the centroid for infrared observations in wavelengths such as 3800 nm, as can be seen from \cite{Descamps_1992}. However in the same paper, the authors obtained a lightcurve observed in Pic du Midi at 800 nm for comparison purposes, but in this wavelength the effect of the volcanoes could not be observed. Since our observations were made in the same wavelength (889 nm), we conclude that these effects can be neglected in our images.

Mutual approximations is a simple, efficient and suitable method for small telescopes. It can be used to continually furnish high precision central instants that can be used to strongly constrain orbit fitting. Ultimately, mutual approximations will significantly contribute to the improvement of the orbits of natural satellites, including the consideration of weak interactions like tidal forces.

\section*{Acknowledgments}
The authors thank the referee Dr. D. Pascu for his constructive comments. BM thanks the financial support by the CAPES/Brazil. MA thanks the CNPq (Grants 473002/2013-2 and 308721/2011-0) and FAPERJ (Grant E-26/111.488/2013). RVM acknowledges the following grants: CNPq-306885/2013, CAPES/Cofecub-2506/2015, FAPERJ/PAPDRJ-45/2013 and  FAPERJ/CNE/05-2015. JIBC acknowledges CNPq for a PQ2 fellowship (process number 308489/2013-6). ADO is thankful for the support of the CAPES (BEX 9110/12-7) FAPERJ/PAPDRJ (E-26/200.464/2015) grants. ARGJ thanks CAPES/Brazil.

\label{lastpage}


\begin{thebibliography}{99}

\bibitem[\protect\citeauthoryear{Arlot et al.}{1982}]{Arlot_1982} Arlot J.-E., Bernard A., Bouchet P. et al., 1982,
A\&A, 111, 151

\bibitem[\protect\citeauthoryear{Arlot et al.}{2012}]{Arlot2012} Arlot J.-E., Emelyanov N. V., Lainey V. et al., 2012,
A\&A, 544, A29

\bibitem[\protect\citeauthoryear{Arlot et al.}{2013}]{Arlot2013} Arlot J.-E., Emelyanov N. V., Aslan Z. et al., 2013,
A\&A, 557, A4

\bibitem[\protect\citeauthoryear{Arlot et al.}{2014}]{Arlot2014} Arlot J.-E., Emelyanov N. V., Varfolomeev M. I. et al., 2014,
A\&A, 572, A120

\bibitem[\protect\citeauthoryear{Assafin et al.}{2009}]{Assafin2009} Assafin M., Vieira-Martins R., Braga-Ribas F. et al., 2009,
The Astronomical Journal, 137, 4046

\bibitem[\protect\citeauthoryear{Assafin et al.}{2011}]{Assafin2011} Assafin M. et al., 2011,
in Gaia follow-up network for the solar system objects :Gaia FUN-SSO
workshop proceedings, held at IMCCE -Paris Observatory, France,
November 29 - December 1, 2010. ISBN 2-910015-63-7, ed. P. Tanga
\& W. Thuillot, 85–88

\bibitem[\protect\citeauthoryear{Butcher \& Stevens}{1981}]{BeS1981} Butcher E. \& Stevens R., 1981,
News Letter Kitt Peak National Observatory, 16, 6

\bibitem[\protect\citeauthoryear{Crida \& Charnoz}{2012}]{Crida_2012} Crida A. \& Charnoz S., 2012,
Science, 338, 1196

\bibitem[\protect\citeauthoryear{De Sitter}{1928}]{Sitter1928} De Sitter W., 1928,
Leiden Ann., 16, 1

\bibitem[\protect\citeauthoryear{Descamps et al.}{1992}]{Descamps_1992} Descamps P., Arlot J.-E., Thuillot W. et al., 1992,
ICARUS, 100, 235

\bibitem[\protect\citeauthoryear{Dias-Oliveira et al.}{2013}]{Diasoliveira2013} Dias-Oliveira A., Vieira-Martins R., Assafin M., et al., 2013,
MNRAS, 432, 225

\bibitem[\protect\citeauthoryear{Emelyanov}{2003}]{Emelyanov2003} Emelyanov N. V., 2003,
Solar System Research, 37(4), 344

\bibitem[\protect\citeauthoryear{Emelyanov}{2009}]{Emelyanov2009} Emelyanov N. V., 2009,
MNRAS, 394, 1037

\bibitem[\protect\citeauthoryear{Emelyanov et al.}{2011}]{Emelyanov2011} Emelyanov N. V., Andreev M. V., Berezhnoi A. A.  et al., 2011,
Solar System Research, 45(3), 264

\bibitem[\protect\citeauthoryear{Faure \& Mensing}{2007}]{Faure2007} Faure G. \& Mensing T. M., 2007, Springer.

\bibitem[\protect\citeauthoryear{Green}{1985}]{Green} Green R. M., 1985, Spherical Astronomy, Cambridge


\bibitem[\protect\citeauthoryear{Harper et al.}{1997}]{Harper1997} Harper D., Murray, C. D., Beurle, K. et al. 1997,
A\&AS, 121, 65

\bibitem[\protect\citeauthoryear{Karkoschka}{1994}]{Karkoschka1994} Karkoschka E., 1994,
ICARUS, 111, 174

\bibitem[\protect\citeauthoryear{Karkoschka}{1998}]{Karkoschka1998} Karkoschka E., 1998,
ICARUS, 133, 134

\bibitem[\protect\citeauthoryear{Kiseleva et al.}{2008}]{Kiseleva2008} Kiseleva T. P., Izmailov I. S., Kiselev A. A. et al., 2008,
Planet. Space Sci., 56, 1908

\bibitem[\protect\citeauthoryear{Lainey et al.}{2004}]{Lainey2004} Lainey V., Arlot J.-E. \& Vienne A., 2004,
A\&A, 427, 371

\bibitem[\protect\citeauthoryear{Lainey et al.}{2009}]{Lainey2009} Lainey V., Arlot J. -E., Karatekin O. \& Van Van Hoolst T., 2009,
Nature, 459, 957

\bibitem[\protect\citeauthoryear{Lieske}{1987}]{Lieske1987} Lieske J. H., 1987,
Astron. Astrophys., 176, 146


\bibitem[\protect\citeauthoryear{Lindegren}{1977}]{Lindegren1977} Lindegren L., 1977,
A\&A, 57, 55

\bibitem[\protect\citeauthoryear{Mason et al.}{1999}]{Mason_1999} Mason B. D., Kaplan G. H., Douglas G. G., et al., 1999,
BAAS, 30, 0509

\bibitem[\protect\citeauthoryear{Pascu}{1994}]{Pascu1994} Pascu D., 1994,
Galactic and Solar System Optical Astrometry. Cambridge Univ. Press, Cambridge, p.304

\bibitem[\protect\citeauthoryear{Peng et al.}{2008}]{Peng2008} Peng Q. Y., Vienne A., Lainey V. \& Noyelles B., 2008,
Planet. Space Sci., 419, 1977

\bibitem[\protect\citeauthoryear{Peng et al.}{2012}]{Peng2012} Peng Q. Y., He F., Lainey V. \& Vienne A., 2012,
MNRAS, 56, 1807

\bibitem[\protect\citeauthoryear{Stone}{1996}]{Stone1996} Stone R. C., 1996,
Publications of the Astronomical Society of the Pacific
, 108, 1051

\bibitem[\protect\citeauthoryear{Stone}{2001}]{Stone2001} Stone R. C., 2001,
The Astronomical Journal, 122, 2723

\bibitem[\protect\citeauthoryear{Veiga et al.}{1987}]{Veiga1987} Veiga C. H., Vieira-Martins, R., Veillet, C. \& Lazzaro, D., 1987,
A\&A, 70, 325

\bibitem[\protect\citeauthoryear{Veiga \& Vieira-Martins}{1994}]{Veiga1994} Veiga C. H. \& Vieira-Martins R., 1994,
A\&A, 107, 551

\bibitem[\protect\citeauthoryear{Veillet \& Ratier}{1980}]{Veillet1980} Veillet C. \& Ratier G., 1980,
A\&A, 89, 342

\bibitem[\protect\citeauthoryear{Vienne et al.}{2001}]{Vienne2001} Vienne, A., Thuillot, W., Veiga, C. H., Arlot, J.-E. \& Vieira-Martins, R., 2001,
A\&A, 380, 727


\end{thebibliography}
\end{document}